\documentstyle[preprint,aps]{revtex}
\begin{document}
\draft
\preprint{UTPT-97-03}
\title{A Self-Organized Critical Universe}
\author{J. W. Moffat}
\address{Department of Physics, University of Toronto,
Toronto, Ontario M5S 1A7, Canada}

\date{\today}
\maketitle

\begin{abstract}%
A model of the universe as a self-organized critical system is considered. The universe 
evolves to a state independently of the initial conditions at the edge of chaos. The critical 
state is an attractor of the dynamics. Random metric fluctuations exhibit noise without 
any characteristic length scales, and the power spectrum for the fluctuations has a self-similar
fractal behavior. In the early universe, the metric fluctuations smear out
the local light cones removing the horizon problem.
\end{abstract}
\pacs{ }

One of the most important problems in modern cosmology concerns the fine-tuning
necessary in the standard cosmology based on general relativity (GR). Why is the
universe so close to being spatially flat after evolving for more than 10 gyr? Why
is it so isotropic and homogeneous? How could such a critical state of the universe
come about without a severe fine tuning of the parameters?

The standard explanation for these questions has been the inflationary
models\cite{Guth}. These models have faced problems that arise mainly from the need
to fine tune certain parameters and initial conditions, e.g., the degree of inhomogeneity
of the initial universe, or in Linde's ``chaotic"  inflation the need
to fine tune parameters at the Planck energy. In the following, we shall study a self-organized
universe which naturally evolves to a critical state without detailed specification of the
initial conditions. The critical state is an {\it attractor} of the system which does not need
to be fine tuned. This is in contrast to the inflationary models which have an
attractor mechanism with critical phase transition points that need to be fine tuned.
In statistical mechanics both kinds of attractor mechanism are known to occur for
physical systems. In contrast to the inflationary paradigm, we shall be concerned with a
self-organized universe in which the spacetime metric fluctuations\cite{Moffat} have a high
degree of cooperative effects at the edge of chaos. Recently, the mixmaster universe has been
shown to be chaotic, i.e., it does not evolve as a self-organized system in
the early universe\cite{Cornish}.

In a recent new approach to gravitational theory\cite{Moffat}, it has been proposed that 
at some length scale much larger than the Planck length, $l_p\approx 10^{-33}$ cm, the
spacetime geometry is fluctuating randomly. In classical GR it is assumed that the spacetime
manifold is $C^2$ smooth down to zero length scales. This seems to be an unacceptably strong
hypothesis considering that known physical systems possess dynamical noise at some
length scale and that cooperative effects are know to occur for many systems proportional
to $V^0$ and not $V^{-a}$ ($a > 0$) where $V$ is a characteristic volume for the physical
system. The metric tensor was treated as a stochastic variable and for a given 
three-geometry ${}^3{\cal G}$, a stochastic differential equation for the momentum conjugate
variable was obtained. A Fokker-Planck equation was derived for the probability density
leading to statistical mechanical predictions for gravitational systems. 

Spatially extended dynamical systems with both temporal and spatial degrees of freedom
are common in biology, physics and chemistry. These systems can evolve with a spatial 
structure that develops as a scale-invariant system exhibiting fractal self-similar
structure.  In statistical mechanics critical phenomena can occur with transition points.
Non-equilibrium systems undergo phase transitions with attractors. But in these dynamical
systems the critical point can be reached only by a parameter fine-tuning and therefore takes
on an accidental description of nature. Inflationary models, as mentioned previously,
fall into this catagory of physical systems. But physical systems exist which evolve as
self-organized critical structures independent of the initial conditions and with no 
fine-tuning of the parameters involved in the system. A well-known example is the
sand-pile model studied by Bak, Tang and Wiesenfeld\cite{Bak}. The sand-pile model
can also be pictured as a system of coupled damped pendula. Energy is dissipated at all
length scales. 

We shall explore our model of the universe using the 
Lema\^{\i}tre-Tolman-Bondi\cite{Lemaitre,Tolman,Bondi,Moffat2}
inhomogeneous, spherically symmetric solution of the GR field equations. The metric
is given by
\begin{equation}
ds^2=dt^2-R^{\prime 2}(r,t)f^{-2}(r)dr^2-R^2(r,t)d\Omega^2,
\end{equation}
where $f$ is an arbitrary function of $r$ only, and the field equations demand that
$R(r,t)$ satisfies
\begin{equation}
2R{\dot R}^2+2R(1-f^2)=F(r)
\end{equation}
with $F$ being an arbitrary function in classical GR of class $C^2, {\dot R}
=\partial R/\partial t$, 
$R^\prime=\partial R/\partial r$, and $d\Omega^2=d\theta^2+\sin^2\theta d\phi^2$.
The universe is filled with dust and there are three solutions, depending on whether
$f^2 < 1, =1, > 1$ and they correspond to elliptic (closed), parabolic (flat), and
hyperbolic (open) cases, respectively.

Let us define the local density parameter, $\Omega(r,t)=\rho(r,t)/\rho_c(r,t)$. There is a
correspondence between the spatial curvature and the sign of $\Omega-1: \Omega-1 < 0,
f^2 > 1$ (open), $\Omega-1=0, f^2=1$ (flat), $\Omega-1 > 0, f^2 < 1$ (closed). The proper
distance between two observers is defined by
\begin{equation}
L_{r\,\rm{prop}}=\int R^\prime(r,t)f^{-1}(r)dr,
\end{equation}
and
\begin{equation}
L_{\perp\,\rm{prop}}=\int R(r,t)d\Omega.
\end{equation}

Let us now turn to a study of our one-dimensional model of the universe. Our model universe has
a large
number of metastable states, which grows exponentially with the expansion of the universe.
These
metastable states correspond to attractors associated with an array of $n$ stable fixed points.
The universe evolves towards a minimally stable state. Let us use a simple model comparable to
the ``sand-pile model"\cite{Bak}. The numbers $\Omega_n$ denote
the coupled values of the density profile associated with the forces of expansion counter-acting
gravity. Then we have
\begin{equation}
\Omega_n=E(n)-E(n+1)
\end{equation}
between successive values of $\Omega$ during the expansion.

The $\Omega_n$ satisfy a nonlinear, discretized diffusion equation with a threshold condition.
We
have
\begin{mathletters}
\begin{eqnarray}
\Omega_n\rightarrow \Omega_n+1,\\
\Omega_{n-1}\rightarrow\Omega_{n-1}-1.
\end{eqnarray}
\end{mathletters}
When $\Omega_n > \Omega_c$, where $\Omega_c$ is a fixed critical value of $\Omega$, a unit
of density profile randomly moves down:
\begin{mathletters}
\begin{eqnarray}
\label{omega1}
\Omega_n\rightarrow\Omega_n-2,\\
\label{omega2}
\Omega_{n\pm 1}\rightarrow\Omega_{n\pm 1}+1,\quad \rm{for}\quad \Omega_n > \Omega_c.
\end{eqnarray}
\end{mathletters}
Eventually all the $\Omega_n$ reach the critical value $\Omega=\Omega_c$, which is the least
stable of the stationary states. Any additional $\Omega_n$ simply falls back from site to site,
falling
off the end $n=N$ and leaving the universe in the minimally stable state. This is the {\it cellular
automaton} model of the universe where the state of the discrete variable $\Omega_n$ at time
$t+1$ depends cooperatively on the state of the variable and its neighbors at time $t$.  The
dynamics leading to the least stable state at some time $t=t_s$ in the expansion is {\it completely
independent of how the universe began.} We can randomly add expansion ``slope",
$\Omega_n\rightarrow \Omega_n+1$ and let the universe satisfy Eqs.(\ref{omega1}) and
(\ref{omega2}). This corresponds to a universe with a random distribution of critical expansion
differences and a uniformly increasing slope of expansion. Starting with a highly unstable
universe
with $\Omega_n > \Omega_c$ for all $n$ and letting the universe contract would lead to the
same minimally stable state.

For our one-dimensional universe, the stable critical state is robust to local perturbations. The 
pressure of expansion will build up to the point {\it where the self-organized critical state is
achieved}.

In a two-dimensional model of the universe, we have $\Omega=\Omega(x,y)$ and\cite{Bak}
\begin{mathletters}
\begin{eqnarray}
\Omega(x-1,y)\rightarrow\Omega(x-1,y)-1,\\
\Omega(x,y-1)\rightarrow\Omega(x,y-1)-1,\\
\Omega(x,y)\rightarrow\Omega(x,y)+2,
\end{eqnarray}
\end{mathletters}
and
\begin{mathletters}
\begin{eqnarray}
\Omega(x,y)\rightarrow\Omega(x,y)-4,\\
\Omega(x,y\pm 1)\rightarrow\Omega(x,y\pm 1)+1,\\
\Omega(x\pm 1,y)\rightarrow\Omega(x\pm 1,y)+1 \quad \rm{for}\quad \Omega(x,y) >
\Omega_c.
\end{eqnarray}
\end{mathletters}
This corresponds to a simple Ising model of the universe with the expansion dynamics involving
next-nearest neighbor interactions.

In two spatial dimensions the minimal stable state is unstable against small fluctuations of the
spacetime geometry and does not describe an attractor of the dynamics. The universe becomes
stable at the point when the minimally stable clusters achieve the level when the fluctuations
cannot be communicated through infinite distances, and there are no characteristic length and
time
scales. The power spectrum for metric fluctuations $S(\Delta g)$ satisfies a power law
\begin{equation}
\label{power}
S(\Delta g)\sim (\Delta g)^{-\beta}
\end{equation}
associated with a self-similar fractal behavior.

We are unable to predict the self-organized critical value, $\Omega=\Omega_c$, without further
dynamical input from GR, but we can explain with this model how the universe has to evolve to
a
stable critical value $\Omega=\Omega_c$, which is completely independent of the initial
conditions and without any fine tuning of parameters. The present universe can be said to be at
the ``edge of chaos" - there is only one possible stable choice for the present expanding universe
whatever its initial configuration. Even if the universe began in a chaotic mix-master 
state\cite{Cornish}, it would soon self-organize itself away from chaos towards the edge of chaos
and a stable, critical state. Observationally $\Omega \leq 1$ today, and this is the only possible
state it can be in according to the fractal scaling of the expansion law and the power law
dependence of the correlation function for the metric fluctuations, Eq. (\ref{power}).

According to our underlying assumption the spacetime geometry fluctuates randomly at some
length scale\cite{Moffat}. If we assume that these fluctuations in the metric are very intense at
the
beginning of the universe, and that they smear out the light cones locally, then for a given short
duration
of time $\Delta t$ after the big-bang there will be communication of information
``instantaneously" throughout the universe which will solve the ``horizon problem", and explain
the present high degree of isotropy and homogeneity of the present universe.

Although the picture of the evolving universe presented here is not incompatible with
inflationary
models, it goes beyond these models by providing a fundamental explanation for the robustness
of
the present universe. In contrast to inflationary models, there is no fine tuning of parameters
required, since the attractor mechanism evolves from a self-organizing dynamical system with a
high degree of cooperative phenomena based on a scale-invariant fractal structure.

Further work must be done to incorporate the diffusion transport equation for $\Omega$ with
GR theory, and thereby provide a basis for more specific dynamical cosmological predictions of
the present state of the universe and the evolution of galaxies.

\acknowledgments

I thank M. A. Clayton and P. Savaria for helpful discussions. This work was supported by the
Natural Sciences
and Engineering Research Council of Canada.

\end{document}